\documentclass[10pt,onecolumn,aps,prd,preprintnumbers,showpacs,superscriptaddress,nofootinbib,amsmath,amssymb,floats,floatfix,showkeys,notitlepage,longbibliography]{revtex4-2}
\usepackage[T1]{fontenc}
\usepackage{orcidlink}
\UseRawInputEncoding
\usepackage{comment}
\usepackage{lipsum}
\usepackage{graphicx}
\usepackage{subfigure}
\UseRawInputEncoding
\usepackage{palatino}
\usepackage{hyperref}
\hypersetup{colorlinks=true,linkcolor=blue,urlcolor=blue,citecolor=blue}
\usepackage[toc,page]{appendix}
\usepackage[normalem]{ulem}
\usepackage{adjustbox}
\usepackage{latexsym}
\usepackage{amsmath}
\usepackage{amssymb}
\usepackage{amsfonts}
\usepackage{commath}
\usepackage{physics}
\usepackage{dcolumn}
\usepackage{bm}
\usepackage{tikz}
\usetikzlibrary{decorations.pathmorphing}
\usepackage{bigints}
\usepackage{array,tabularx,multirow,booktabs}
\usepackage[tracking=true]{microtype}
\microtypesetup{expansion=false}
\usepackage{soul} %for highlighting
\SetTracking{}{500}
\SetTracking{encoding={*}, shape=sc}{40}
\UseRawInputEncoding %for inputenc error%
\allowdisplaybreaks

\usepackage[utf8]{inputenc}
\usepackage{xcolor} % For colored text if needed

\begin{document} \sloppy
\title{Fixed ADM reconstruction of extended uncertainty principle black holes}

\author{Nikko John Leo S. Lobos \orcidlink{0000-0001-6976-8462}}
\email{nslobos@ust.edu.ph}
\affiliation{Electronics Engineering Department, University of Santo Tomas, Espa\~na Boulevard, Sampaloc, Manila 1008, Philippines}

\author{Ali \"Ovg\"un \orcidlink{0000-0002-9889-342X}}
\email{ali.ovgun@emu.edu.tr}
\affiliation{Physics Department, Faculty of Arts and Sciences, Eastern Mediterranean University, Famagusta, 99628 North Cyprus via Mersin 10, Turkiye.}

\author{Reggie C. Pantig \orcidlink{0000-0002-3101-8591}} 
\email{rcpantig@mapua.edu.ph}
\affiliation{Physics Department, School of Foundational Studies and Education, Map\'ua University, 658 Muralla St., Intramuros, Manila 1002, Philippines.}

%\date{\today}

\begin{abstract}
We study a phenomenological fixed-ADM prescription for an extended-uncertainty-principle (EUP) correction to the Schwarzschild exterior. Rather than treat mass constancy as a consequence of the EUP, we hold the charge measured at spatial infinity fixed so that we can compare exteriors at the same physical mass. We impose the EUP-shifted horizon radius and Hawking temperature and select the slowest admissible inverse-power representative. Its Einstein tensor defines an effective anisotropic tensor, but we do not infer an independent matter model, quantum state, or perturbation theory from that definition. For a positive EUP parameter, the effective energy density is negative throughout the exterior, while the transverse null and strong energy conditions fail over an extended radial region. The area entropy and the formal entropy obtained from the reduced relation $\mathrm{d}M=T\,\mathrm{d}S$ disagree at leading order; we interpret this mismatch as evidence that the thermodynamic description lacks work terms or an action-based entropy law, not as a definite EUP entropy correction. An explicit one-parameter family with the same mass, horizon, and temperature shows that the orbit, lensing, and test-field eikonal shifts depend on the chosen representative. Our results therefore characterize the minimal member of a phenomenological family rather than universal consequences of the EUP.
\end{abstract}


\maketitle

\section{Introduction} \label{sec1}
Black hole thermodynamics is one of the few domains in which general relativity, quantum field theory, and statistical mechanics meet in a quantitatively rigid way. Bekenstein's identification of black hole entropy with horizon area gave the first indication that gravitational dynamics encodes thermodynamic information \cite{Bekenstein:1973ur}. Hawking's derivation of black hole radiation then fixed the temperature in terms of the surface gravity and converted the analogy into a semiclassical prediction \cite{Hawking:1975vcx}. These results are universal enough to survive many changes of microscopic interpretation, yet restrictive enough that any proposed quantum correction to a black hole metric must confront a basic consistency problem: the corrected geometry, the corrected horizon, and the corrected temperature cannot be assigned independently.

Modified uncertainty principles provide a phenomenological route into this problem. The generalized uncertainty principle was originally motivated by the expectation that localization at trans-Planckian energies should be obstructed by gravitational backreaction \cite{Maggiore:1993rv}. Algebraic realizations of minimal-length quantum mechanics made this idea technically precise by deforming the canonical commutator and studying its Hilbert-space representation \cite{Kempf:1994su}. Micro-black hole gedanken experiments gave an independent derivation of a minimal resolvable length from the interplay between Heisenberg localization and Schwarzschild collapse \cite{Scardigli:1999jh}. In black hole physics, this line of reasoning led naturally to GUP-corrected temperatures, modified evaporation laws, and remnant scenarios \cite{Adler:2001vs}. The common structural feature is that the uncertainty relation supplies thermodynamic observables before it supplies a spacetime geometry.

The extended uncertainty principle is complementary. Instead of introducing a short-distance correction controlled by the Planck length, it introduces a large-distance correction associated with background curvature, cosmological scales, or infrared gravitational structure. In the context of de Sitter and anti-de Sitter black holes, the EUP reproduces the qualitative form of curvature-dependent thermodynamic corrections when the horizon scale is used as the localization length \cite{Bolen:2004sq}. Mignemi subsequently showed that an EUP-type relation can be obtained from the geometry and symmetry algebra of constant-curvature spacetime, clarifying that the EUP is not merely a dimensional analogy with the GUP but may encode infrared geometric information \cite{Mignemi:2009ji}. These observations make the EUP attractive for black hole phenomenology, especially when one is interested in corrections that become appreciable for large black holes rather than Planckian remnants.

A particularly direct implementation was proposed by Mureika, who used the EUP to construct an effective Schwarzschild metric with a corrected horizon radius, temperature, photon sphere, ISCO, and weak-field potential \cite{Mureika:2018gxl}. The construction is economical: the EUP modifies the characteristic momentum scale, the modified momentum is translated into an effective mass, and this effective mass is inserted into the Schwarzschild lapse. This produces explicit phenomenological predictions and has the practical advantage of preserving the familiar one-function Schwarzschild form. However, this economy also exposes a conceptual limitation. If the correction is absorbed into a mass parameter multiplying the asymptotic Newtonian tail, the resulting metric is locally Schwarzschild outside the horizon with a shifted mass. The correction then does not represent a genuine fixed-ADM deformation of the exterior geometry; it is a mass renormalization unless an independent prescription identifies the bare mass rather than the asymptotic mass as the physical observable. Subsequent work has developed this EUP-inspired black hole program in several directions. EUP corrections have been studied in black hole temperature and Unruh-effect calculations, in Schwarzschild and Reissner--Nordstr\"om thermodynamics, in higher-order EUP thermodynamic models, and in phase-transition analyses \cite{Chung:2019iwp,Hassanabadi:2021kyv,Hamil:2021ilv,Okcu:2022iwl}. The same effective-metric idea has also been used to compute weak and strong lensing observables, finite-distance deflection effects, shadows in dark-matter environments, generalized extended-uncertainty-principle black holes, and direct astrophysical bounds on the EUP length scale \cite{Lu:2019wfi,Kumaran:2019qqp,Pantig:2021zqe,Lobos:2022jsz,Okcu:2022sio}. More recent developments have further connected EUP-type relations with horizon thermodynamics in cosmological settings, relativistic curvature-induced uncertainty relations, rotating G\"odel backgrounds, spacetime-curvature induced uncertainty principles, and nonlocal black hole models \cite{Dabrowski:2019wjk,Wagner:2021thc,Pantig:2025kxm,Pantig:2024asu,Capozziello:2025iwn}. These works show that the EUP literature is no longer limited to a single Schwarzschild substitution rule, but they also sharpen the question addressed here: whether the EUP correction can be reconstructed as a genuine fixed-ADM exterior deformation rather than as an asymptotic mass relabeling.

The fixed-ADM choice defines a comparison scheme, not a physical necessity imposed by the EUP. In asymptotically flat gravity, the ADM mass is the charge measured at spatial infinity, so holding it fixed lets us isolate changes in the exterior at equal mass. An EUP model could instead alter the total energy, binding energy, or asymptotic gravitational charge; in that case an ADM-mass shift would belong to the physical effect. We do not exclude that possibility. We ask the narrower conditional question of how the exterior may change when the observer at infinity assigns the same mass before and after the EUP correction.

A term that changes the leading inverse-radius coefficient changes the mass read at infinity. Under our comparison scheme, the correction must therefore enter through faster-decaying metric functions, a redshift factor, an effective source, modified field equations, or nonlocal boundary observables. Recent studies of heuristic effective metrics emphasize that horizon and temperature inputs do not select a unique radial profile and that finite series prescriptions may yield misleading conclusions \cite{Ong:2023jkp}. We therefore pose an inverse problem and identify one minimal representative while retaining the residual freedom explicitly.

We apply this inverse problem to a static, spherically symmetric Schwarzschild exterior. We treat the EUP horizon and temperature as boundary inputs, hold the asymptotic charge fixed, and choose the slowest admissible inverse-power member. We then define an effective anisotropic tensor from the Einstein tensor. The Bianchi identity guarantees its covariant conservation, but it neither identifies the degrees of freedom that produce it nor establishes their stability.

We do not claim a microscopic EUP source. We impose asymptotic flatness, fixed ADM mass, the EUP horizon displacement, and the EUP temperature at leading order. These requirements admit infinitely many radial profiles. The simplest member supplies correlated, model-dependent shifts in the effective density and pressures, curvature scalars, circular-orbit quantities, weak deflection, and suitable test-field eikonal frequencies. None of these shifts provides an independent consistency test or a universal EUP prediction without additional field dynamics.

The paper proceeds as follows. Section \ref{sec2} states the fixed-ADM inverse problem, selects the minimal metric, exhibits its residual freedom, and outlines the fixed-charge Kerr extension. Section \ref{sec3} examines the associated effective tensor, energy conditions, and curvature invariants. Section \ref{sec4} separates the area law from the formal reduced first-law integral. Section \ref{sec5} studies circular geodesics, and Sec. \ref{sec6} treats weak deflection and test-field eikonal frequencies. We use units with $c=\hbar=k_B=1$, keep $G$ explicit, adopt the mostly-plus signature, and identify the mass in the asymptotic Newtonian tail with the ADM mass.

\section{Fixed-ADM reconstruction from EUP horizon observables} \label{sec2}
\subsection{Inverse observables and obstruction to pure mass renormalization}
Let $M$ denote the ADM mass measured at spatial infinity and define the Schwarzschild radius $r_s=2GM$. The dimensionless EUP control parameter is taken to be
\begin{equation}
\lambda=\frac{\alpha r_s^2}{L_*^2}
       =\frac{4\alpha G^2M^2}{L_*^2},
\qquad 
|\lambda|\ll 1 .
\label{1.1}
\end{equation}
The EUP-corrected horizon and temperature observables to be reconstructed are
\begin{equation}
r_+=r_s(1+\lambda)+\mathcal{O}(\lambda^2),
\qquad
T_+=\frac{1}{4\pi r_s}(1-\lambda)+\mathcal{O}(\lambda^2).
\label{1.2}
\end{equation}
Our comparison prescription keeps $M$ equal to the ADM mass. The coefficient of the $1/r$ term therefore remains unchanged. This restriction isolates a fixed-mass exterior deformation; it does not follow from the EUP itself and does not rule out models in which the EUP changes the ADM charge.

We use the static, spherically symmetric areal-gauge ansatz
\begin{equation}
\mathrm{d}s^2=-\mathrm{e}^{2\psi(r)}F(r)\,\mathrm{d}t^2+\frac{\mathrm{d}r^2}{F(r)}+r^2\mathrm{d}\Omega^2 .
\label{1.3}
\end{equation}
At first order in $\lambda$ we write
\begin{equation}
F(r)=1-\frac{r_s}{r}+\lambda f(r)+\mathcal{O}(\lambda^2),
\qquad
\psi(r)=\lambda \chi(r)+\mathcal{O}(\lambda^2).
\label{1.4}
\end{equation}
Fixed ADM mass, asymptotic flatness, and the absence of an EUP-induced Newtonian-tail shift impose \cite{Szabados:2009eka}
\begin{equation}
f(r)=o(r^{-1}),
\qquad
\chi(r)=o(r^{-1})
\qquad
(r\rightarrow \infty).
\label{1.5}
\end{equation}
The Landau little-$o$ notation indicates that the EUP correction must decay faster than the Newtonian tail. The condition $F(r_+)=0$ with $r_+=r_s+\lambda r_s+\mathcal{O}(\lambda^2)$ gives
\begin{equation}
f(r_s)=-1 .
\label{1.6}
\end{equation}
The surface gravity of the metric in Eq. \eqref{1.3} is
\begin{equation}
\kappa=\frac12 \mathrm{e}^{\psi(r_+)}F'(r_+).
\label{1.7}
\end{equation}
Expanding Eq. \eqref{1.7} about $r=r_s$ and demanding $\kappa=(2r_s)^{-1}(1-\lambda)+\mathcal{O}(\lambda^2)$ yields the second reconstruction condition
\begin{equation}
\chi(r_s)+r_s f'(r_s)=1 .
\label{1.8}
\end{equation}

\subsection{Minimal fixed-ADM solution}
Within an integer inverse-power expansion in $r_s/r$, the slowest admissible falloff compatible with fixed ADM mass is $r^{-2}$. Equation \eqref{1.5} alone does not exclude logarithmic or noninteger-power corrections that satisfy $o(r^{-1})$. We therefore take the $r^{-2}$ term as the minimal representative only within this restricted expansion class.
\begin{equation}
f(r)=-\frac{r_s^2}{r^2},
\qquad
\chi(r)=-\frac{r_s^2}{r^2}.
\label{1.9}
\end{equation}
It satisfies Eq. \eqref{1.6}, since $f(r_s)=-1$, and Eq. \eqref{1.8}, since $r_s f'(r_s)=2$ and $\chi(r_s)=-1$. The reconstructed EUP geometry is consequently
\begin{equation}
F(r)=1-\frac{r_s}{r}-\lambda\frac{r_s^2}{r^2}+\mathcal{O}(\lambda^2),
\qquad
\psi(r)=-\lambda\frac{r_s^2}{r^2}+\mathcal{O}(\lambda^2).
\label{1.10}
\end{equation}
Equivalently, the redshift function $A(r)=\mathrm{e}^{2\psi(r)}F(r)$ is
\begin{equation}
A(r)=1-\frac{r_s}{r}
+\lambda\left(
-\frac{3r_s^2}{r^2}
+\frac{2r_s^3}{r^3}
\right)
+\mathcal{O}(\lambda^2).
\label{1.11}
\end{equation}
The usual mass-renormalized EUP metric corresponds instead to
\begin{equation}
F_{\rm mr}(r)=1-\frac{r_s(1+\lambda)}{r}.
\label{1.12}
\end{equation}
Equation \eqref{1.12} reproduces the EUP horizon shift only by changing the coefficient of the $1/r$ tail. If the physical ADM radius is denoted by $r_{\rm ADM}=r_s(1+\lambda)$, then Eq. \eqref{1.12} is exactly Schwarzschild with radius $r_{\rm ADM}$. Equation \eqref{1.10} is therefore our minimal representative within the chosen fixed-mass comparison scheme, rather than a solution selected uniquely by the EUP.

\subsection{Residual freedom and sensitivity of the shifts}

The two horizon conditions constrain only values and derivatives at $r=r_s$; they do not fix the radial profiles. To display the remaining freedom, let $x=r_s/r$ and introduce a dimensionless constant $\eta$ of order unity. The family
\begin{equation}
f_\eta(r)=-x^2+\eta\left(x^2-x^3\right),
\qquad
\chi_\eta(r)=-(1+\eta)x^2
\label{1.13}
\end{equation}
decays faster than $1/r$ and obeys $f_\eta(r_s)=-1$ together with $\chi_\eta(r_s)+r_sf_\eta'(r_s)=1$ for every $\eta$. Hence every member has the same ADM mass, horizon displacement, and temperature through first order. The choice in Eq. \eqref{1.9} is the member $\eta=0$.

Direct expansion gives
\begin{equation}
\begin{aligned}
\frac{r_{\rm ph}}{r_s}
&=\frac32+\frac{2(8+\eta)}{9}\lambda+\mathcal{O}(\lambda^2),
&
\frac{b_{\rm ph}}{r_s}
&=\frac{3\sqrt3}{2}\left[1+\frac{2(5+\eta)}{9}\lambda\right]+\mathcal{O}(\lambda^2),
\\
\frac{r_{\rm ISCO}}{r_s}
&=3+\frac{43+8\eta}{9}\lambda+\mathcal{O}(\lambda^2),
&
r_s\Omega_{\rm ph}
&=\frac{2}{3\sqrt3}\left[1-\frac{2(5+\eta)}{9}\lambda\right]+\mathcal{O}(\lambda^2),
\\
r_s\Lambda_{\rm ph}
&=\frac{2}{3\sqrt3}\left[1-\frac{2(11+\eta)}{27}\lambda\right]+\mathcal{O}(\lambda^2),
&
\widehat\alpha
&=\frac{2r_s}{b}+\left[\frac{15\pi}{16}+\frac{\pi(7+\eta)}{4}\lambda\right]\frac{r_s^2}{b^2}
+\mathcal{O}\!\left(\frac{r_s^3}{b^3},\lambda^2\right).
\end{aligned}
\label{1.14}
\end{equation}
The $\eta$ dependence proves that these shifts are properties of the selected radial representative. We use $\eta=0$ in the remaining sections, and we do not present its numerical coefficients as universal EUP predictions.

\subsection{Kerr extension at fixed asymptotic charges}
\label{sec_kerr-roadmap}

The fixed-ADM comparison has a natural rotating analogue, but it cannot be obtained by inserting a spin parameter into Eq. \eqref{1.10}. The reference geometry must be Kerr with both ADM mass $M$ and angular momentum $J$ held fixed. In an asymptotically Cartesian mass-centered gauge, these charges are fixed by preserving the leading terms
\begin{align}
g_{tt}=-1+\frac{2GM}{r}+\mathcal{O}(r^{-2}),
\qquad
g_{t\phi}=-\frac{2GJ}{r}\sin^2\theta+\mathcal{O}(r^{-2}).    
\label{1.15}
\end{align}

A suitable perturbative starting point is therefore
\begin{align}
g_{\mu\nu}=g^{\rm Kerr}_{\mu\nu}(M,J)
+\lambda h_{\mu\nu}(r,\theta)+\mathcal{O}(\lambda^2),
\qquad
\chi_\ast=\frac{J}{GM^2},\quad |\chi_\ast|<1,
\label{1.16}
\end{align}
where $h_{\mu\nu}$ is stationary, axisymmetric, and asymptotically charge preserving. Its nonrotating limit must reproduce Eq. \eqref{1.10}. Rotation changes the inverse problem qualitatively: even after equatorial symmetry and circularity are imposed, the independent metric functions depend on both $r$ and $\theta$, rather than on $r$ alone.

The horizon data also require a new prescription. A regular rotating horizon is generated by $\xi_H=\partial_t+\Omega_H\partial_\phi$; the conditions $\xi_H^2=0$, constant surface gravity $\kappa_H=2\pi T_H$, and constant angular velocity $\Omega_H$ must hold on the same Killing horizon. The Schwarzschild EUP input in Eq. \eqref{1.2} specifies neither the invariant rotating localization scale nor the correction to $\Omega_H$. One tentative choice is the areal scale $R_H=\sqrt{A_H/(4\pi)}$, which reduces to the horizon radius in the nonrotating limit, and to use the EUP relation to prescribe $A_H$ and $\kappa_H$. An additional dynamical or thermodynamic condition is then needed to determine $\Omega_H$. Even if all three horizon quantities are supplied, they constrain only boundary values of several functions and do not select a unique exterior. A direct Newman--Janis transformation can therefore provide at most a candidate ansatz; its ADM charges, horizon regularity, constancy of $\kappa_H$ and $\Omega_H$, nonrotating limit, and effective source must be checked independently.

The minimum practical extension is to work to first order in $\lambda$ about exact subextremal Kerr while retaining arbitrary $\chi_\ast$. In a fixed Boyer--Lindquist--type gauge, one may expand $h_{\mu\nu}$ in equatorially even angular harmonics, retain the slowest charge-preserving inverse powers, and impose the asymptotic, horizon, and Schwarzschild-limit conditions above. If an action or source model is supplied, its linearized field equations must determine the remaining functions. Without such dynamics, the unfixed coefficients should be retained as sensitivity parameters, as $\eta$ is in Eq. \eqref{1.13}, rather than hidden by a particular rotating ansatz. A slow-rotation expansion provides a useful check at small $\chi_\ast$, but it is not sufficient for comparison with rapidly rotating systems. Only after this background problem is fixed should one calculate the inclination-dependent photon region, shadow, ISCO, lensing, or test-field eikonal modes; the resulting EUP shifts must then be separated from changes in spin, inclination, and the emission model. This program identifies the conditions required for an observational Kerr extension without claiming that the present static reconstruction already determines one.

\section{Effective Einstein tensor and phenomenological source} \label{sec3}
\subsection{Einstein tensor of the selected geometry}
Let
\begin{equation}
\Phi(r)=\psi(r)+\frac12\ln F(r),
\qquad
A(r)=\mathrm{e}^{2\Phi(r)} .
\label{2.1}
\end{equation}
For the ansatz in Eq. \eqref{1.3}, the nonzero independent mixed Einstein components are
\begin{equation}
\begin{aligned}
G^t{}_{t}
&=-\frac{1-F-rF'}{r^2},
\\
G^r{}_{r}
&=\frac{F-1+rF'}{r^2}+\frac{2F\psi'}{r},
\\
G^\theta{}_{\theta}
&=G^\phi{}_{\phi}
=F\!\left(\Phi''+\Phi'^2+\frac{\Phi'}{r}\right)
+\frac{F'\Phi'}{2}
+\frac{F'}{2r}.
\end{aligned}
\label{2.2}
\end{equation}
Substitution of Eq. \eqref{1.10}, followed by expansion through $\mathcal{O}(\lambda)$, gives
\begin{equation}
\begin{aligned}
G^t{}_{t}
&=\lambda\frac{r_s^2}{r^4}+\mathcal{O}(\lambda^2),
\\
G^r{}_{r}
&=\lambda\frac{r_s^2}{r^4}
\left(5-\frac{4r_s}{r}\right)
+\mathcal{O}(\lambda^2),
\\
G^\theta{}_{\theta}
&=G^\phi{}_{\phi}
=\lambda\frac{r_s^2}{r^4}
\left(-5+\frac{7r_s}{r}\right)
+\mathcal{O}(\lambda^2).
\end{aligned}
\label{2.3}
\end{equation}
After selecting the metric, we define an associated effective tensor through the ordinary Einstein equations,
\begin{equation}
T^\mu{}_{\nu,{\rm eff}}
=\frac{1}{8\pi G}G^\mu{}_{\nu}[g].
\label{2.4}
\end{equation}
This relation is a definition, not an independent derivation of matter degrees of freedom. In the anisotropic-fluid notation $T^\mu{}_\nu={\rm diag}(-\rho,p_r,p_\perp,p_\perp)$,
\begin{equation}
\begin{aligned}
\rho
&=-\frac{\lambda r_s^2}{8\pi G r^4}
+\mathcal{O}(\lambda^2),
\\
p_r
&=\frac{\lambda r_s^2}{8\pi G r^4}
\left(5-\frac{4r_s}{r}\right)
+\mathcal{O}(\lambda^2),
\\
p_\perp
&=\frac{\lambda r_s^2}{8\pi G r^4}
\left(-5+\frac{7r_s}{r}\right)
+\mathcal{O}(\lambda^2).
\end{aligned}
\label{2.5}
\end{equation}
The Bianchi identity guarantees conservation of the tensor defined in Eq. \eqref{2.4}, which one may check explicitly as
\begin{equation}
p_r'
+(\rho+p_r)\Phi'
+\frac{2}{r}(p_r-p_\perp)
=\mathcal{O}(\lambda^2).
\label{2.6}
\end{equation}
Using the zeroth-order $\Phi'=\frac12\partial_r\ln(1-r_s/r)$ in Eq. \eqref{2.6}, the left-hand side vanishes identically through $\mathcal{O}(\lambda)$. Conservation alone does not supply a matter action, an equation of state, a quantum state, dynamical stability, or perturbation equations. We therefore use $T^\mu{}_{\nu,{\rm eff}}$ only as a phenomenological description of the selected background.

\subsection{Energy conditions and curvature invariants}
For $\lambda>0$, the effective tensor has negative energy density outside the horizon. The relevant null combinations are
\begin{equation}
\begin{aligned}
\rho+p_r
&=\frac{\lambda r_s^2}{8\pi G r^4}
4\left(1-\frac{r_s}{r}\right)
+\mathcal{O}(\lambda^2),
\\
\rho+p_\perp
&=\frac{\lambda r_s^2}{8\pi G r^4}
\left(-6+\frac{7r_s}{r}\right)
+\mathcal{O}(\lambda^2).
\end{aligned}
\label{2.7}
\end{equation}
The radial null condition holds for $r\geq r_s$, whereas the transverse null condition fails for $r>7r_s/6$. Because $\rho<0$, both the weak and dominant energy conditions fail throughout the exterior. Moreover, $\rho+p_r+2p_\perp=2K(-3+5r_s/r)+\mathcal{O}(\lambda^2)$, where $K=\lambda r_s^2/(8\pi G r^4)>0$. The strong energy condition therefore holds only in the near-horizon interval $r_s\leq r\leq7r_s/6$ at this order and fails beyond it. At large radius, $\rho=-K+\mathcal{O}(r^{-5})$, $p_r=5K+\mathcal{O}(r^{-5})$, and $p_\perp=-5K+\mathcal{O}(r^{-5})$. Thus the negative-energy, highly anisotropic tail reaches spatial infinity, although it decays as $r^{-4}$. In terms of the mass function $F=1-2Gm(r)/r$, one has $m(r)=M+\lambda r_s^2/(2Gr)+\mathcal{O}(\lambda^2)$ and $m(\infty)-m(r_+)=-\lambda M+\mathcal{O}(\lambda^2)$. This extended violation is a property of the chosen metric, not evidence that a realizable semiclassical EUP state exists.

The Ricci scalar and Kretschmann scalar are
\begin{equation}
R
=
\lambda\frac{r_s^2}{r^4}
\left(4-\frac{10r_s}{r}\right)
+\mathcal{O}(\lambda^2),
\label{2.8}
\end{equation}
and
\begin{equation}
R_{\alpha\beta\gamma\delta}R^{\alpha\beta\gamma\delta}
=
\frac{12r_s^2}{r^6}
+\lambda\left(
\frac{112r_s^3}{r^7}
-\frac{88r_s^4}{r^8}
\right)
+\mathcal{O}(\lambda^2).
\label{2.9}
\end{equation}
Thus the exterior is not Ricci-flat. The change cannot be removed by relabeling the Schwarzschild mass, but the nonzero curvature does not identify the microscopic physics that supports it.

\section{Horizon thermodynamics} \label{sec4}
\subsection{Horizon radius and surface gravity}
The horizon is the largest root of $F(r)=0$. For the minimal reconstructed metric,
\begin{equation}
r_+
=
\frac{r_s}{2}\left(1+\sqrt{1+4\lambda}\right)
=
r_s(1+\lambda)+\mathcal{O}(\lambda^2).
\label{3.1}
\end{equation}
The derivative of $F$ is
\begin{equation}
F'(r)=\frac{r_s}{r^2}
+\frac{2\lambda r_s^2}{r^3}
+\mathcal{O}(\lambda^2).
\label{3.2}
\end{equation}
Using Eqs. \eqref{1.7}, \eqref{3.1}, and \eqref{3.2}, the surface gravity is
\begin{equation}
\kappa
=
\frac{1}{2r_s}(1-\lambda)+\mathcal{O}(\lambda^2).
\label{3.3}
\end{equation}
The Hawking temperature is therefore
\begin{equation}
T_+
=
\frac{\kappa}{2\pi}
=
\frac{1}{4\pi r_s}(1-\lambda)+\mathcal{O}(\lambda^2)
=
\frac{1}{8\pi GM}
\left(
1-\frac{4\alpha G^2M^2}{L_*^2}
\right)
+\mathcal{O}(L_*^{-4}).
\label{3.4}
\end{equation}
This reproduces the EUP temperature while preserving the ADM mass.

\subsection{Area law versus first-law entropy}

The horizon area is
\begin{equation}
A_+=4\pi r_+^2
=
4\pi r_s^2(1+2\lambda)+\mathcal{O}(\lambda^2).
\label{3.5}
\end{equation}
If we interpret Eq. \eqref{2.4} as Einstein gravity coupled to an effective matter sector, the stationary gravitational entropy remains the area entropy,
\begin{equation}
S_{\rm area}
=
\frac{A_+}{4G}
=
\frac{\pi r_s^2}{G}(1+2\lambda)+\mathcal{O}(\lambda^2).
\label{3.6}
\end{equation}
For comparison only, if we suppress every matter or work contribution and integrate the reduced relation $\mathrm{d}M=T_+\mathrm{d}S$, we obtain the formal quantity
\begin{equation}
S_{\rm FL}(M)
=
\int^M\frac{\mathrm{d}M'}{T_+(M')}
=
4\pi GM^2
+\frac{8\pi\alpha G^3M^4}{L_*^2}
+\mathcal{O}(L_*^{-4}).
\label{3.7}
\end{equation}
In terms of $r_s$, this formal quantity is
\begin{equation}
S_{\rm FL}
=
\frac{\pi r_s^2}{G}
\left(1+\frac{\lambda}{2}\right)
+\mathcal{O}(\lambda^2).
\label{3.8}
\end{equation}
Equations \eqref{3.6} and \eqref{3.8} are inequivalent. If one combines the area expression with the same reduced relation, the required temperature would be
\begin{equation}
T_{\rm area}
=
\frac{1}{4\pi r_s}(1-4\lambda)+\mathcal{O}(\lambda^2),
\label{3.9}
\end{equation}
which differs from the EUP temperature in Eq. \eqref{3.4}. We do not infer a non-area entropy from this mismatch. Along the family of metrics, the effective tensor also varies, so the complete first law may contain an additional work term that the reduced relation omits. If the tensor instead represents modified gravitational dynamics, both the first law and entropy must follow from a specified action, for example through the covariant phase-space and Noether-charge method \cite{Iyer:1994ys}. The mismatch therefore shows that the present thermodynamic description is incomplete; $S_{\rm FL}$ is only the result of the stated reduced integral.

\begin{figure}
\centering
\includegraphics[width=0.48\textwidth]{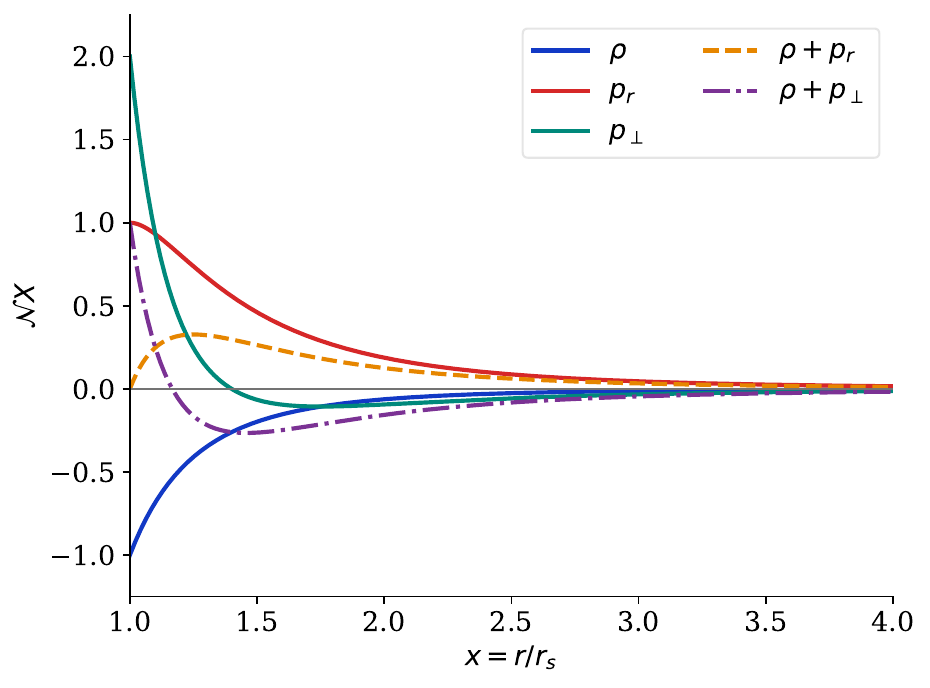}
\includegraphics[width=0.48\textwidth]{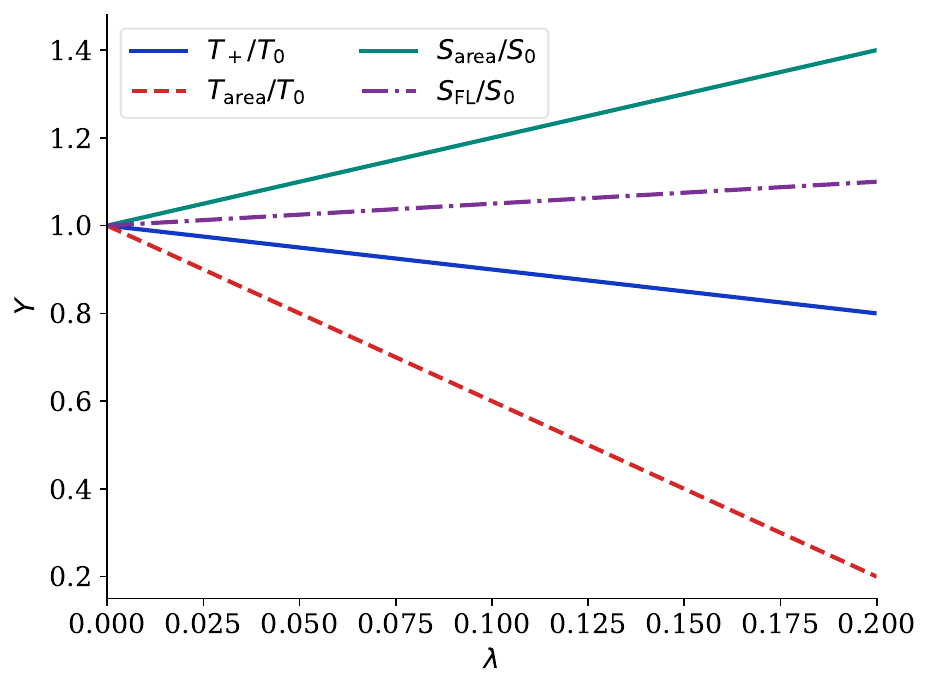}
\caption{The left panel shows the normalized effective-tensor and energy-condition profiles as functions of $x=r/r_s$.
The vertical variable is $\mathcal{N}X$, where $\mathcal{N}=8\pi G r_s^2/\lambda$ and
$X\in\{\rho,p_r,p_\perp,\rho+p_r,\rho+p_\perp\}$. The right panel compares the normalized temperature and entropy relations in the minimal fixed-ADM model. Its horizontal variable is $\lambda$, and its vertical variable is $Y$.
The plotted normalized quantities are $T_+/T_0$, $T_{\rm area}/T_0$, $S_{\rm area}/S_0$, and $S_{\rm FL}/S_0$, with
$T_0=(4\pi r_s)^{-1}$ and $S_0=\pi r_s^2/G$.}
\label{fig_source-profile}
\end{figure}

The left panel of Fig. \ref{fig_source-profile} shows the extended negative-energy and anisotropic profile described above. The radial null combination is nonnegative for $x\geq1$, while the transverse combination changes sign at $x=7/6$ and remains negative farther out. The right panel visualizes the algebraic mismatch: $T_+/T_0=1-\lambda$, $T_{\rm area}/T_0=1-4\lambda$, $S_{\rm area}/S_0=1+2\lambda$, and $S_{\rm FL}/S_0=1+\lambda/2$. These curves compare mutually incomplete thermodynamic prescriptions; they do not select the physical entropy without a complete first law or an action.

\subsection{Heat capacity}
At fixed $L_*$ and $\alpha$, Eq. \eqref{3.4} gives
\begin{equation}
T_+(M)
=
\frac{1}{8\pi G}
\left(
\frac{1}{M}
-\frac{4\alpha G^2M}{L_*^2}
\right)
+\mathcal{O}(L_*^{-4}).
\label{3.10}
\end{equation}
The heat capacity is therefore
\begin{equation}
C
=
\frac{\mathrm{d}M}{\mathrm{d}T_+}
=
-8\pi GM^2
\left(1-\lambda\right)
+\mathcal{O}(\lambda^2).
\label{3.11}
\end{equation}
Along the stated one-parameter path, $C$ remains negative for small positive $\lambda$ and its magnitude decreases relative to Schwarzschild. This derivative alone does not establish thermodynamic stability because the effective sector has no specified work variables or susceptibilities.

\section{Geodesic observables} \label{sec5}
Every result below refers to the minimal member $\eta=0$. The quantities derive from the same assumed metric and are therefore correlated signatures of that member, not separate tests of its physical viability. Equation \eqref{1.14} displays how residual radial freedom changes representative examples.

\subsection{Null and timelike effective potentials}
Restricting to the equatorial plane, the conserved energy and angular momentum are
\begin{equation}
E=A(r)\dot t,
\qquad
L=r^2\dot\phi,
\qquad
A(r)=\mathrm{e}^{2\psi(r)}F(r).
\label{4.1}
\end{equation}
For null geodesics $\delta=0$ and for timelike geodesics $\delta=1$, the radial equation is
\begin{equation}
\dot r^2
=
F(r)\left[
\frac{E^2}{A(r)}
-\frac{L^2}{r^2}
-\delta
\right].
\label{4.2}
\end{equation}
The redshift function required in the circular-orbit analysis is
\begin{equation}
A(r)
=
1-\frac{r_s}{r}
+\lambda\left(
-\frac{3r_s^2}{r^2}
+\frac{2r_s^3}{r^3}
\right)
+\mathcal{O}(\lambda^2).
\label{4.3}
\end{equation}

\subsection{Photon sphere and shadow radius}
The photon-sphere and shadow observables are obtained from the standard null circular-orbit construction for static spherical metrics \cite{Perlick:2021aok,Berry:2020ntz}. For instance, the photon sphere is determined by extremizing $A(r)/r^2$,
\begin{equation}
rA'(r)-2A(r)=0 .
\label{4.4}
\end{equation}
Solving Eq. \eqref{4.4} perturbatively gives
\begin{equation}
r_{\rm ph}
=
\frac{3r_s}{2}
+\frac{16}{9}\lambda r_s
+\mathcal{O}(\lambda^2).
\label{4.5}
\end{equation}
The critical impact parameter is
\begin{equation}
b_{\rm ph}
=
\frac{r_{\rm ph}}{\sqrt{A(r_{\rm ph})}}
=
\frac{3\sqrt{3}}{2}r_s
\left(
1+\frac{10}{9}\lambda
\right)
+\mathcal{O}(\lambda^2).
\label{4.6}
\end{equation}
Thus the shadow size increases for $\lambda>0$ at fixed ADM mass.

\subsection{ISCO observables}
For a timelike circular orbit, the conserved quantities are \cite{Berry:2020ntz}
\begin{equation}
L^2(r)
=
\frac{r^3A'(r)}{2A(r)-rA'(r)},
\qquad
E^2(r)
=
\frac{2A(r)^2}{2A(r)-rA'(r)}.
\label{4.7}
\end{equation}
The ISCO is fixed by $\mathrm{d}L^2/\mathrm{d}r=0$. Using Eq. \eqref{4.3}, one obtains
\begin{equation}
r_{\rm ISCO}
=
3r_s+\frac{43}{9}\lambda r_s
+\mathcal{O}(\lambda^2).
\label{4.8}
\end{equation}
The associated energy, angular momentum, and orbital frequency are
\begin{equation}
\begin{aligned}
E_{\rm ISCO}
&=
\frac{2\sqrt{2}}{3}
\left(
1+\frac{5}{54}\lambda
\right)
+\mathcal{O}(\lambda^2),
\\
L_{\rm ISCO}
&=
\sqrt{3}\,r_s
\left(
1+\frac{31}{27}\lambda
\right)
+\mathcal{O}(\lambda^2),
\\
\Omega_{\rm ISCO}
&=
\frac{1}{3\sqrt{6}\,r_s}
\left(
1-\frac{31}{18}\lambda
\right)
+\mathcal{O}(\lambda^2).
\end{aligned}
\label{4.9}
\end{equation}
The outward displacement of the ISCO and the decrease of $\Omega_{\rm ISCO}$ are both stronger than what follows from a pure Schwarzschild mass relabeling at the same parameter value.

\section{Weak deflection and eikonal ringing} \label{sec6}
We again set $\eta=0$. The weak-deflection and frequency shifts characterize this member; they do not follow from the horizon and temperature inputs alone.

\subsection{Weak deflection angle}
Let $b$ be the impact parameter and $r_0$ the distance of closest approach. They are related by
\begin{equation}
b^2=\frac{r_0^2}{A(r_0)}.
\label{5.1}
\end{equation}
The bending angle for null geodesics is \cite{Ono:2019hkw}
\begin{equation}
\hat\alpha
=
2\int_{r_0}^{\infty}
\frac{b\,\mathrm{d}r}
{r^2\sqrt{F(r)\left[A(r)^{-1}-b^2r^{-2}\right]}}
-\pi .
\label{5.2}
\end{equation}
With $y=r_0/r$ and $p=r_s/r_0$, Eq. \eqref{5.2} becomes
\begin{equation}
\hat\alpha
=
2\int_0^1
\frac{\mathrm{d}y}
{\sqrt{
F(r_0/y)
\left[
A(r_0)/A(r_0/y)-y^2
\right]
}}
-\pi .
\label{5.3}
\end{equation}
Expanding the integrand to $\mathcal{O}(p^2)$ and $\mathcal{O}(\lambda)$ gives
\begin{equation}
\int_0^1
\frac{\mathrm{d}y}
{\sqrt{
F(r_0/y)
\left[
A(r_0)/A(r_0/y)-y^2
\right]
}}
=
\frac{\pi}{2}
+p
+\left(
\frac{15\pi}{32}
-\frac12
+\frac{7\pi}{8}\lambda
\right)p^2
+\mathcal{O}(p^3,\lambda^2).
\label{5.4}
\end{equation}
Using $p=r_s/b+\mathcal{O}(r_s^2/b^2)$, the bending angle in terms of the invariant impact parameter is
\begin{equation}
\hat\alpha
=
\frac{2r_s}{b}
+
\left(
\frac{15\pi}{16}
+\frac{7\pi}{4}\lambda
\right)
\frac{r_s^2}{b^2}
+O\!\left(\frac{r_s^3}{b^3},\lambda^2\right).
\label{5.5}
\end{equation}
The first term is the usual Schwarzschild deflection. The EUP correction enters at second post-Minkowskian order because fixed ADM mass forbids a correction to the $1/r$ potential.

\begin{figure}
\centering
\includegraphics[width=0.60\textwidth]{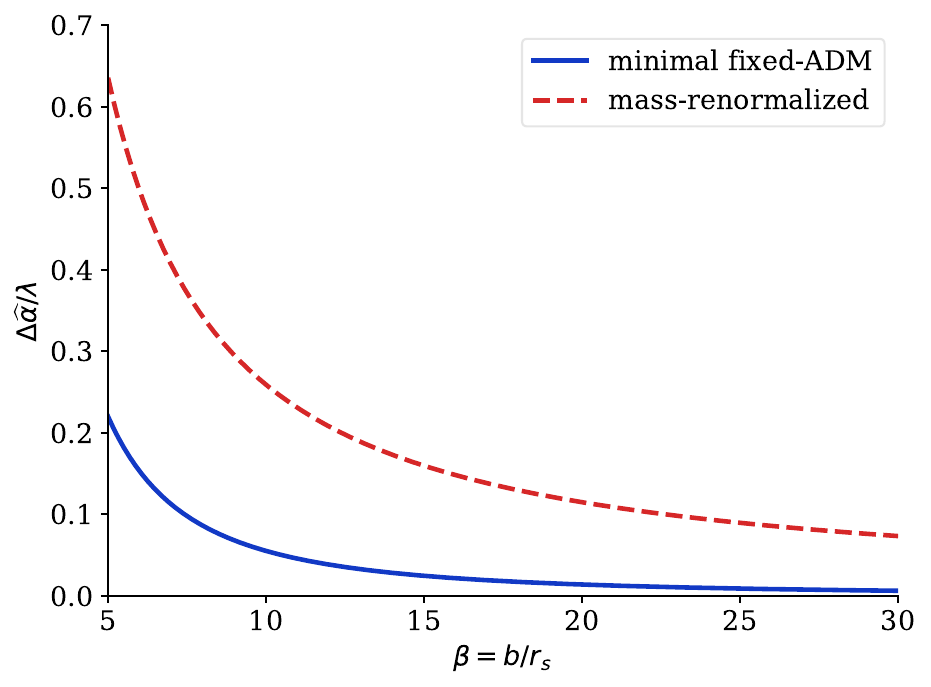}
\caption{Weak-deflection residual hierarchy as a function of $\beta=b/r_s$.
The vertical variable is $\Delta\hat{\alpha}/\lambda$.
The plotted curves are
$\Delta\hat{\alpha}_{\rm F}/\lambda=(7\pi/4)\beta^{-2}$
for the fixed-ADM reconstruction and
$\Delta\hat{\alpha}_{\rm mr}/\lambda=2\beta^{-1}+(15\pi/8)\beta^{-2}$
for the mass-renormalized Schwarzschild comparison at fixed parameter $r_s$.}
\label{fig_deflection-residual}
\end{figure}

Figure \ref{fig_deflection-residual} shows the different post-Minkowskian hierarchy of the EUP correction in the two constructions. In the fixed-ADM geometry, the absence of a correction to the asymptotic Newtonian tail removes the $\beta^{-1}$ contribution, so the EUP residual begins at order $\beta^{-2}$. By contrast, the mass-renormalized Schwarzschild comparison contains a leading $\beta^{-1}$ residual because the EUP parameter changes the coefficient of the inverse-radius potential. The separation of the two curves therefore gives a direct analytic visualization of the fixed-ADM principle: preserving the ADM charge suppresses the EUP lensing correction by one post-Minkowskian order relative to a pure mass relabeling.

\subsection{Test-field eikonal frequencies}
We restrict this subsection to suitable minimally coupled test fields in the eikonal limit. We do not compute gravitational quasinormal modes of the effective system. Such modes require perturbation equations for both the metric and the degrees of freedom behind $T^\mu{}_{\nu,{\rm eff}}$; the background tensor in Eq. \eqref{2.4} does not supply those equations. The null-orbit correspondence may fail for coupled gravitational sectors, as discussed in Ref. \cite{Konoplya:2017wot}.

The angular frequency of the unstable null orbit is
\begin{equation}
\Omega_{\rm ph}
=
\frac{\sqrt{A(r_{\rm ph})}}{r_{\rm ph}}
=
\frac{2}{3\sqrt{3}\,r_s}
\left(
1-\frac{10}{9}\lambda
\right)
+\mathcal{O}(\lambda^2).
\label{5.6}
\end{equation}
The coordinate-time Lyapunov exponent is
\begin{equation}
\Lambda_{\rm ph}^2
=
-\frac{r_{\rm ph}^2}{2B(r_{\rm ph})}
\left.
\frac{\mathrm{d}^2}{\mathrm{d}r^2}
\left(
\frac{A(r)}{r^2}
\right)
\right|_{r=r_{\rm ph}},
\qquad
B(r)=\frac{1}{F(r)}.
\label{5.7}
\end{equation}
For the reconstructed metric,
\begin{equation}
\Lambda_{\rm ph}
=
\frac{2}{3\sqrt{3}\,r_s}
\left(
1-\frac{22}{27}\lambda
\right)
+\mathcal{O}(\lambda^2).
\label{5.8}
\end{equation}
For the stated test-field regime, the eikonal spectrum is \cite{Cardoso:2008bp,Konoplya:2017wot}
\begin{equation}
\omega_{\ell n}
=
\left(\ell+\frac12\right)\Omega_{\rm ph}
-\mathrm{i}\left(n+\frac12\right)\Lambda_{\rm ph}
+\mathcal{O}(\ell^{-1}).
\label{5.9}
\end{equation}
or explicitly,
\begin{equation}
\omega_{\ell n}
=
\left(\ell+\frac12\right)
\frac{2}{3\sqrt{3}\,r_s}
\left(
1-\frac{10}{9}\lambda
\right)
-\mathrm{i}\left(n+\frac12\right)
\frac{2}{3\sqrt{3}\,r_s}
\left(
1-\frac{22}{27}\lambda
\right)
+\mathcal{O}(\lambda^2,\ell^{-1}).
\label{5.10}
\end{equation}

\subsection{Observable distinction from mass renormalization}
For the mass-renormalized geometry $F_{\rm mr}=1-r_s(1+\lambda)/r$, all strong-field radii are obtained by the replacement $r_s\mapsto r_s(1+\lambda)$. At fixed parameter $r_s$, this gives
\begin{equation}
\begin{array}{c|c|c}
\text{Quantity}
&
\text{Mass-renormalized Schwarzschild}
&
\text{Fixed-ADM reconstructed geometry}
\\[2mm]
\hline
r_+
&
r_s(1+\lambda)
&
r_s(1+\lambda)
\\[1mm]
r_{\rm ph}
&
\dfrac{3r_s}{2}(1+\lambda)
&
\dfrac{3r_s}{2}+\dfrac{16}{9}\lambda r_s
\\[3mm]
b_{\rm ph}
&
\dfrac{3\sqrt{3}}{2}r_s(1+\lambda)
&
\dfrac{3\sqrt{3}}{2}r_s
\left(1+\dfrac{10}{9}\lambda\right)
\\[3mm]
r_{\rm ISCO}
&
3r_s(1+\lambda)
&
3r_s+\dfrac{43}{9}\lambda r_s
\end{array}
\label{5.11}
\end{equation}
If instead one compares at fixed physical ADM mass, the mass-renormalized geometry has no invariant deviation from Schwarzschild. The minimal fixed-ADM member remains distinguishable through its Ricci tensor, circular geodesics, shadow radius, second-post-Minkowskian bending, and test-field eikonal damping. The sizes of these shifts change when one varies the admissible radial profile.

\begin{figure}
\centering
\includegraphics[width=0.60\textwidth]{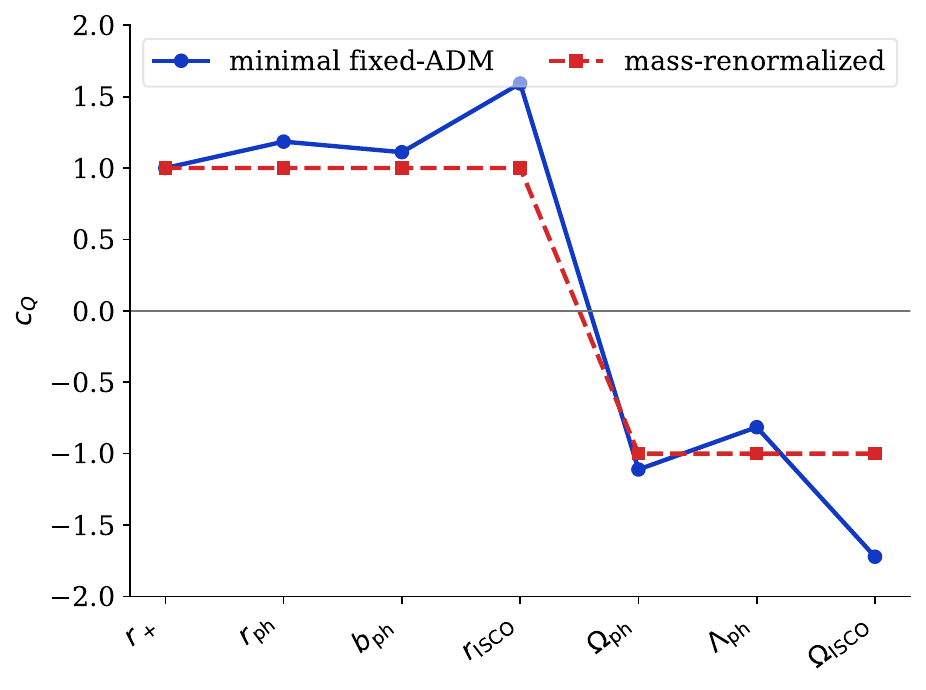}
\caption{
Linear fractional-shift spectrum for fixed-ADM and mass-renormalized EUP geometries. The vertical variable is
$c_Q=\left.\mathrm{d}\ln(Q/Q_{\rm Schw})/\mathrm{d}\lambda\right|_{\lambda=0}$, where $Q$ denotes the observable on the horizontal axis.
The solid curve with circular markers denotes the minimal fixed-ADM member, while the dashed curve with square markers denotes the mass-renormalized Schwarzschild comparison at fixed parameter $r_s$.}
\label{fig_shift-spectrum}
\end{figure}

Figure \ref{fig_shift-spectrum} displays the first-order response for the minimal member $\eta=0$. At fixed parameter $r_s$, it differs from the mass-renormalized Schwarzschild comparison for every exterior quantity shown. The plot summarizes Eq. \eqref{5.11}; Eq. \eqref{1.14} separately shows that the fixed-ADM coefficients vary across admissible profiles.

\section{Conclusion} \label{sec7}
We have examined a conditional fixed-ADM scheme for an EUP-inspired Schwarzschild exterior. Holding the mass measured at infinity fixed lets us compare geometries at equal asymptotic charge, but the EUP does not require this choice. Horizon and temperature inputs leave residual radial freedom. Our $r^{-2}$ ansatz is the simplest member, and the explicit $\eta$ family shows that its observable coefficients are model dependent.

We define the effective tensor from the Einstein tensor after choosing the metric. Its Bianchi conservation ensures background consistency but does not provide an action, equation of state, quantum state, stability result, or perturbation law. For positive $\lambda$, the weak and dominant energy conditions fail throughout the exterior, while the transverse null and strong conditions fail beyond $7r_s/6$. We therefore regard the extended negative-energy atmosphere as a phenomenological property of the minimal metric, not as a demonstrated semiclassical EUP source.

We also interpret the entropy mismatch more narrowly. In Einstein gravity, the area expression remains the gravitational entropy and the varying effective sector may add a work term; in modified gravity, an action must determine the entropy. The reduced integral supplies no unique physical entropy by itself. Likewise, the photon sphere, shadow, ISCO, weak bending, and test-field eikonal frequencies are correlated consequences of one metric. We make no claim about gravitational quasinormal modes without coupled perturbation equations.

Future work should derive the exterior from an explicit action or quantum state, classify the full set of admissible radial profiles, and formulate their coupled perturbations. A rotating extension also requires more than a direct Newman--Janis step: one must fix both asymptotic mass and angular momentum, satisfy horizon temperature and angular-velocity conditions, handle additional metric functions, and derive a conserved axisymmetric effective tensor. These tasks will determine which, if any, of the present shifts survive a more complete physical theory.

\acknowledgments
R. P. and A. \"O.  would like to acknowledge networking support of the COST Action CA21106 - COSMIC WISPers in the Dark Universe: Theory, astrophysics and experiments (CosmicWISPers), the COST Action CA22113 - Fundamental challenges in theoretical physics (THEORY-CHALLENGES), the COST Action CA21136 - Addressing observational tensions in cosmology with systematics and fundamental physics (CosmoVerse), the COST Action CA23130 - Bridging high and low energies in search of quantum gravity (BridgeQG), and the COST Action CA23115 - Relativistic Quantum Information (RQI) funded by COST (European Cooperation in Science and Technology). A. \"O. also thanks to EMU, TUBITAK, ULAKBIM (Turkiye) and SCOAP3 (Switzerland) for their support.

\bibliography{ref}

\end{document}